\documentclass{aastex63}
\usepackage{textcomp}

\usepackage{graphicx}
\usepackage{subfigure}
\usepackage{epsfig}
\usepackage{rotating}
\usepackage{floatrow}

\DeclareFloatFont{scriptsize}{\scriptsize}
\shortauthors{Lin et al.}

\usepackage{amssymb}

\begin{document}

\title{A compact object with a K-type star companion in the solar neighborhood: a wide postcommon envelope binary with a white dwarf candidate}

\correspondingauthor{Jie Lin}
\email{j.lin@pku.edu.cn}

\author{Jie Lin}
\affiliation{Department of Astronomy, Xiamen University, Xiamen, Fujian 361005, People’s Republic of China}

\author{Hailiang Chen}
\affiliation{Yunnan Observatories, Chinese Academy of Sciences, Kunming 650216, People’s Republic of China}
\affiliation{International Centre of Supernovae, Yunnan Key Laboratory, Kunming 650216, P. R. China}

\author{Bojun Wang}
\affiliation{National Astronomical Observatory, Chinese Academy of Sciences, Beijing 100101, P.~R.~China}

\author{Yudong Luo}
\affiliation{School of Physics, Peking University, and Kavli Institute for Astronomy and Astrophysics, Peking University, Beijing, 100871, China}

\author{Wenshi Tang}
\affiliation{Department of Astronomy, Xiamen University, Xiamen, Fujian 361005, People’s Republic of China}

\author{Bo Huang}
\affiliation{Institut de Ci\`{e}ncies de l'Espai (ICE-CSIC), Campus UAB, Can Magrans S/N, E-08193 Cerdanyola del Vall\`{e}s, Catalonia, Spain}


\begin{abstract}
Post-common envelope binaries (PCEBs) consisting of a white dwarf (WD) plus a main-sequence (MS) star can constrain current prescriptions of common envelope evolution (CEE) and calibrate theoretical models of binary formation and evolution. Most PCEBs studied to date have typical orbital periods of hours to a few days and can be well explained by assuming inefficient CEE to expel the envelope. However, there are currently several systems with relatively wide orbital periods ($>$18\,days). To explain these wide PCEBs, additional sources of energy have been suggested to be taken into account. Here, we present the discovery and observational characterization of a compact object ($M\,\geq\,0.58\,\rm M_{\odot}$) with a K-type star companion in the solar neighborhood ($d\sim 112$\,pc) and an orbital period of $P_{\rm orb}\sim 14$\,days. The compact object binary is likely to be a system consisting
of a WD and a barium dwarf, making it the shortest-period barium star binary candidate. Such a system with an orbital period within the gap between tight and wide binaries provides a test of whether additional energy sources are required to explain its formation. Using binary evolution models, we investigate the evolutionary history of this wide PCEB system and find that the observed properties of this source can be explained without invoking any extra energy source. 
\end{abstract}


\section{Introduction}

Common envelope evolution (CEE) is one of the major unsolved problems in modern astronomy and astrophysics. CEE is formed when a giant donor star overflows its Roche lobe, typically as a consequence of dynamically unstable mass transfer. During CEE, angular momentum and orbital energy are extracted from the orbit to eject the envelope. If the envelope is successfully ejected, a close binary containing a compact object and its main-sequence (MS) companion will be left. 
If the envelope is not successfully ejected, the CEE will lead to the merger of the binary system. Modeling of CEE is an important progress toward the understanding of the formation of a wide variety of compact binary systems, such as cataclysmic variables \citep{1976IAUS...73...75P,1995cvs..book.....W}, X-ray binaries \citep{1998ApJ...493..351K}, double-degenerate binaries \citep{1984ApJ...277..355W}, binary neutron stars (NSs; \citealt{1991PhR...203....1B}), and binary black holes (BHs; \citealt{2016Natur.534..512B}). Due to the complex physical processes involved in the CEE, simplified models of CEE based on energy or angular momentum conservation are commonly used in terms of modeling the evolution of large numbers of binaries and understanding the possible formation pathways of individual systems. 

In current binary evolution models, CEE is commonly approximated by a parameterized energy equation, i.e. a fraction of the liberated orbital energy, known as the common-envelope (CE) efficiency ($\alpha_{\rm CE}$), is equal to the binding energy of the envelope to determine the post-CEE orbital separation \citep{1976IAUS...73...75P,1995cvs..book.....W,1986ApJ...311..742I,1993PASP..105.1373I}. This model has been calibrated by comparing the predicted binary populations with the abundance of observed post-common-envelope binaries (PCEBs) that contain white dwarfs (WDs). Previous modeling attempts to reconstruct the CE phase for observed post-common envelope binaries (PCEBs) and empirically constrain $\alpha_{\rm CE}$. Several calculations have found that the vast majority of systems, with typical orbital periods of hours to a few days, can be best reproduced by assuming inefficient CE evolution with $\alpha_{\rm CE} \sim 0.25$ \citep{2010A&A...520A..86Z}. However, it remains unclear whether additional energy sources play an important role in unbinding the envelope for wide PCEBs. 
One notable example is IK Peg \citep{1993MNRAS.262..277W}, a 22 days binary containing a massive WD ($1.19\,\rm M_{\odot}$), whose orbit can only be explained when additional energy sources, such as the recombination energy of hydrogen and helium released in the expanding envelope of the WD progenitor, are included in the energy budget \citep{2010MNRAS.403..179D,2010A&A...520A..86Z}.
In addition,  several wide WD + MS binaries with orbital periods from a few tens to a few hundreds of days have been discovered via the Kepler survey and Gaia Data Release 3 (DR3) catalog\citep{2014Sci...344..275K,2018AJ....155..144K,2024MNRAS.52711719Y}. These long-period PCEBs, such as IK Peg, cannot be explained by classical CE evolution without contributions from additional energy sources in the CE energy balance. However, the formation of long-period PCEBs may remain if mass transfer begins during a thermally pulsing asymptotic giant branch (TP-AGB) phase of the donor star \citep{2024A&A...687A..12B}. According to this formation
scenario, the slow neutron-capture ({\it s}-process) elements were transferred
to the main-sequence stars from a more evolved companion when
the latter was in its TP-AGB phase \citep{1980ApJ...238L..35M}. This can result in an overabundance
of barium and other {\it s}-process elements on the surface of the
MS stars, which are historically known as Barium stars.

Large spectroscopic surveys, such as the Large Sky Area Multi-Object Fiber Spectroscopic Telescope (LAMOST) and the Gaia mission, have observed millions of stars and built extensive spectroscopic databases. Radial-velocity modulation detected in optical spectra offers a promising method for identifying unseen compact objects in binary systems with stellar companions. These databases open a new opportunity to search for BHs, NSs, and long-period PCEBs binaries with massive WD \citep{2020ApJ...905...38R,2022ApJ...940..165Y,2022NatAs...6.1203Y,2022ApJ...936...33Z,El-Badrya,El-Badryb,2023ApJ...944L...4L,2023AJ....165..187Q,2024ApJ...964..101Z,2024A&A...686L...2G,2024ApJ...969..114L,2024MNRAS.52711719Y}.

In this article, we report a single-lined binary consisting of a K-type MS star and a compact object, which is likely a long-orbital-period PCEB composed of a WD and a barium dwarf. The paper is organized as follows. The identification of a wide PCEB with a WD candidate from the catalog of Gaia DR3 single-lined spectroscopic binaries (SB1s) is described in Section 2. The data analysis and the result are presented in Section 3. Section 4 discusses the possibilities for the nature of the unseen companion and presents models of WD progenitors and constraints on CEE. Our conclusions are presented in Section 5.

\section{Discovery}
Gaia DR3 provides orbital solutions for more than 181000 single-lined spectroscopic binaries (SB1s), enabling the identification of candidate systems that may host compact object companions such as WDs, NSs, or BHs \citep{2023A&A...674A..34G}. Furthermore, the Fourth Fermi Large Area Telescope Source Catalog (4FGL) DR3 contains more than 2000 unassociated $\gamma$-ray sources, representing potential reservoirs harboring a significant population of undiscovered spider pulsar binary systems (e.g., \citealt{2024ApJ...960L...5L}). In this project, we performed a crossmatch between Gaia DR3 SB1s and the 4FGL-DR3 catalog, selecting systems exhibiting significantly large mass functions in their Gaia DR3 radial-velocity (RV) solutions to search for spider pulsar binary system candidates. We discovered a candidate with the optical source of Gaia DR3 located at R.A. (J2000) = $04^{\rm h}47^{\rm m}28^{\rm s}.48$ and Decl. (J2000) = $+24^\circ{} 44^{'}52.^{''}4$, inside the 95 $\%$ error region of the $\gamma-$ray source 4FGL J0447.2+2446 (see Fig 1).

In Gaia DR3, the source$\_$id of J0447 is 147167226196998272 with a bright ($G$=11.95) K-type MS star. Its DR3 parallax of $\omega_{\rm DR3}= 8.96\pm 0.04 $\,mas implies a distance of $d \sim$ 112 pc. The RV solution of Gaia DR3 gives the orbital period ($P_{\rm orb}=13.396 \pm 0.002$ days), eccentricity ($e=0.03 \pm 0.01$ ), semiamplitude ( $K=42.37 \pm 0.53$\,$\rm km$\,$\rm s^{-1}$) and center-of-mass velocity ($\gamma = 59.20 \pm 0.36$\,$\rm km$\,$\rm s^{-1}$). For $\gamma$-ray source 4FGL J0447.2+2446, its $\gamma$-ray spectrum can be fitted with a logParabola model \citep{2022ApJS..260...53A}. 4FGL J0447.2+2446 has a 0.1$-$100\,GeV flux of (4.19$\pm$0.73)\,$\times10^{-12}$$\rm erg$\,$\rm s^{-1}$$\rm cm^{-2}$, corresponding to a luminosity of $L_{\gamma}=(6.3 \pm 1.1)\times10^{30}$\,$\rm erg$\,$\rm s^{-1}$ at a fiducial distance of 112 pc. This estimated $\gamma$-ray luminosity of J\,0447 is far less than within the range of luminosities ($4\times10^{32}$ - $4\times10^{34}$\,$\rm erg$\,$\rm s^{-1}$) of the spider pulsar family. 
Notably, PSR J0447+2447 is spatially coincident within the 68 $\%$ confidence level with the $\gamma$-ray source 4FGL J0447.2+2446 (see Fig 1). This suggests that the Gaia source is likely a chance alignment with the gamma-ray source, which has a different more likely counterpart.
PSR J0447+2447 is an isolated millisecond pulsar (MSP) with a spin period of 2.99 ms and a dispersion measure (DM) of 32.907\,$\rm cm^{-3}$\,pc \citep{2023MNRAS.522.5152W}. For PSR J0447 + 2447, the distance of 0.92 kpc derived from the YMW16 model \citep{2017ApJ...835...29Y} corresponds to a gamma-ray luminosity of $L_{\gamma}=(4.28 \pm 0.75)\times10^{32}$\,$\rm erg$\,$\rm s^{-1}$, which lies within the characteristic range for millisecond pulsars. Therefore, 4FGL J0447.2+2446 is  essentially the $\gamma$-ray counterpart of the PSR J0447+2447 rather than J\,0447. However, the possibility that the observed $\gamma$-ray flux originates from multiple sources cannot be entirely excluded, since two counterparts lie within the 95$\%$ confidence level. Hence, multiwavelength observations are required to determine the nature of the invisible star in J0447.

\begin{figure}[htp]
\centering
\includegraphics[width=15cm]{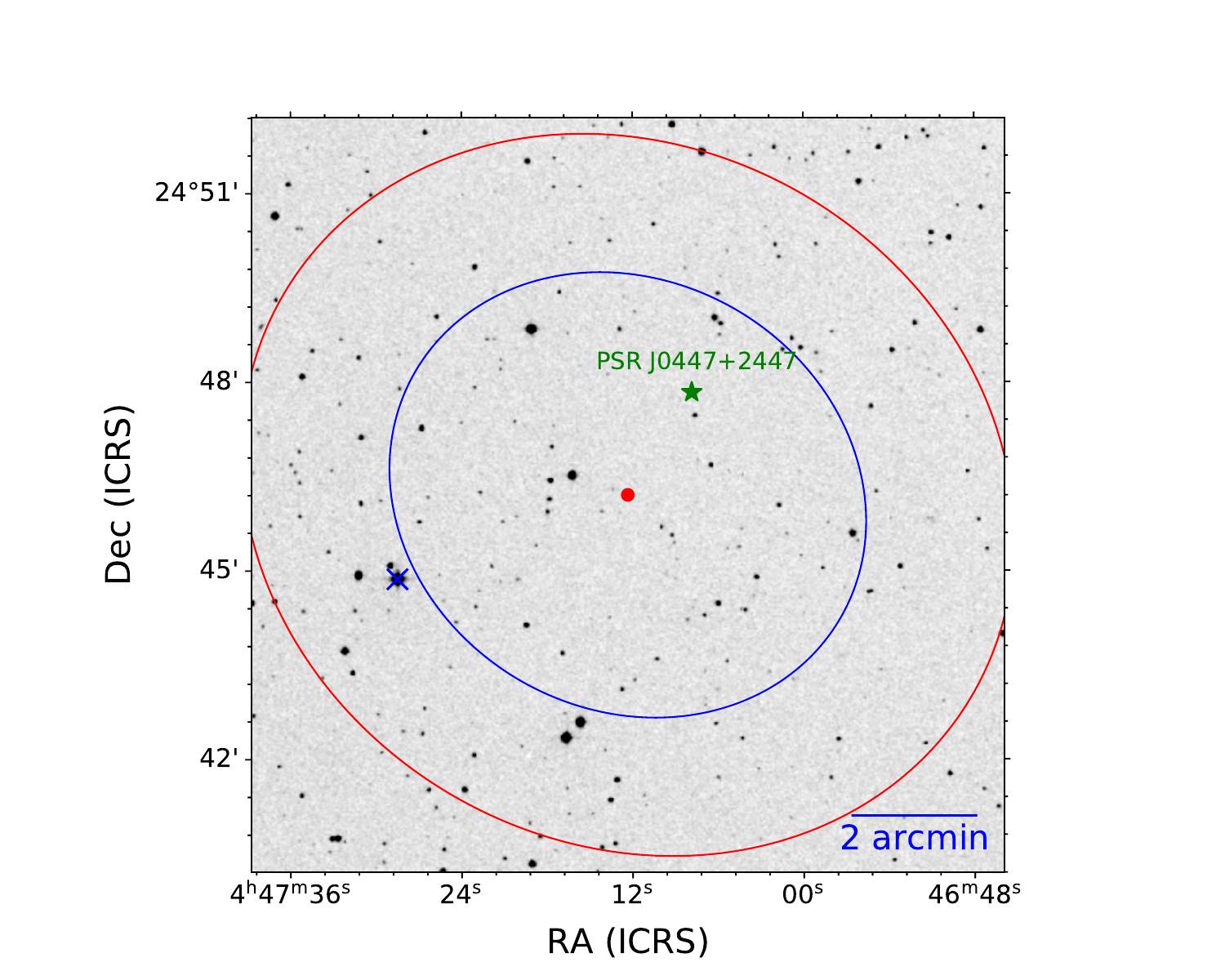}
\caption{The Digitized Sky Survey (DSS) image of the 4FGL J0447.2+2446. The red and blue ellipses show $68\%$ error ellipse and $95\%$ error ellipse of 4FGL catalog, respectively. The PSR J0447+2447 (green star) is localized within the 68\% confidence level of 4FGL J0447.2+2446, while the optical source (blue cross) position is contained in the 95\% confidence region. }
\end{figure}

\section{Data analysis and results}

\begin{figure}[htp]
\centering
\includegraphics[width=15cm]{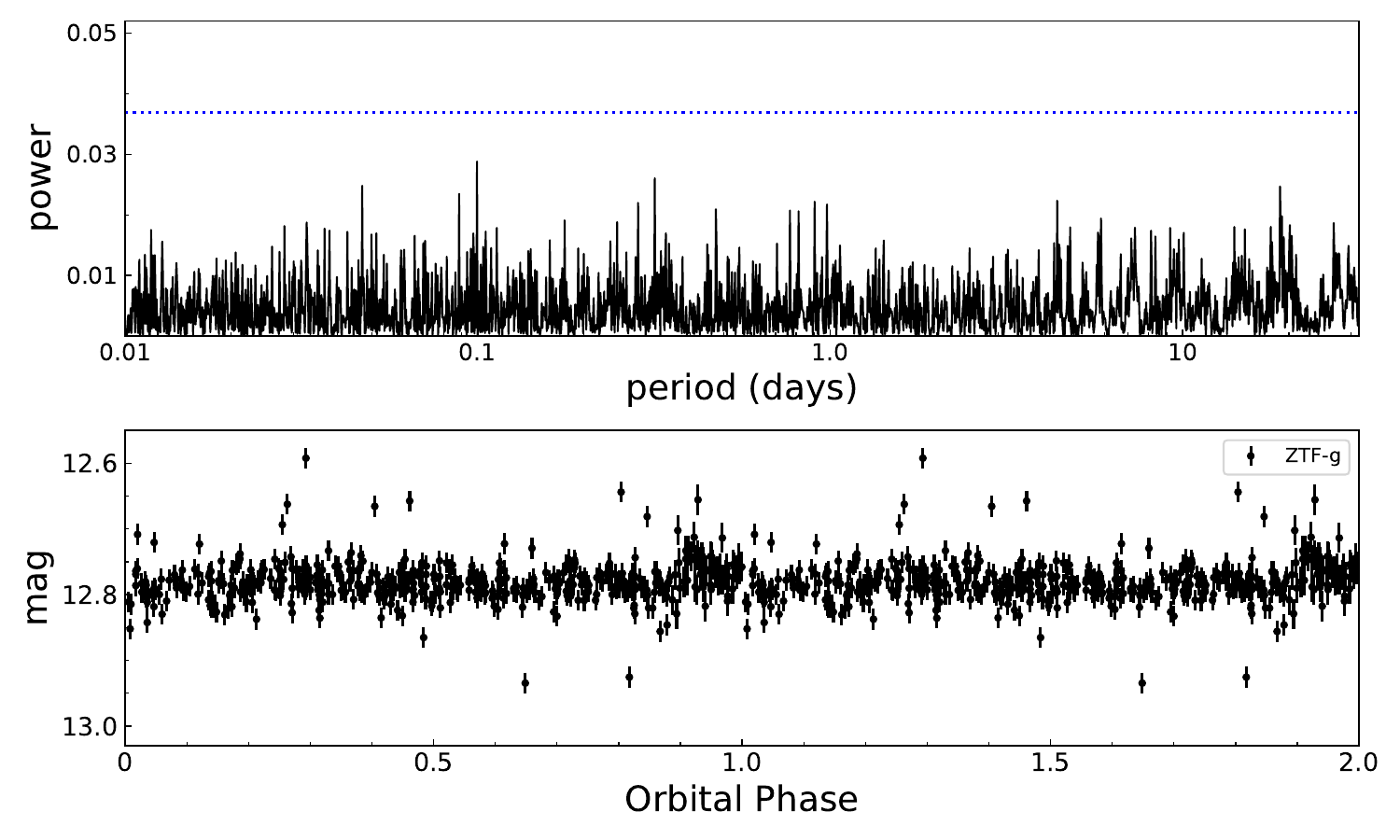}
\caption{The top panel shows the Lomb-Scargle periodogram of the complete ZTF dataset (MJD 58204-60609), with dashed horizontal line representing the $95\%$ confidence limits. The bottom panel shows the ZTF light curves folded at the orbital phase derived from LAMOST RV measurements.}
\end{figure}

\subsection{Optical Photometry}	
The Zwicky Transient Facility (ZTF) DR23 archive\footnote{https://irsa.ipac.caltech.edu/cgi-bin/Gator/nph-dd} contains about five hundred brightness measurements of J0447 in the g band covering about 6.6 yr (MJD 58204–60609). We use the Lomb-Scargle method \citet{1976Ap&SS..39..447L,1982ApJ...263..835S}, to search for periodic brightness variations in the range of 0.01-30 days. The resulting periodogram is shown in the upper panel of Fig. 2. No periodic signals exceeding the 2 $\sigma$ confidence level were detected in the periodogram. The blue dashed lines in the plot represent the 2 $\sigma$ confidence level.

In addition, we retrieved photometry of the source in Gaia DR3 ($G$, $G_{\rm BP}$, and $G_{\rm RP}$), Two Micron All Sky Survey ({\it J}, {\it H} and {\it K$_{\rm s}$}), Sloan Digital Sky Survey (SDSS; $u$), Wide-field Infrared
Survey Explorer (W1 and W2) and TESS (T), and 
fit the spectral energy distribution (SED) with the spectrAl eneRgy dIstribution bAyesian moDel averagiNg fittEr \citep[astroARIADNE \footnote{https://github.com/jvines/astroARIADNE};][]{2022MNRAS.513.2719V}. The SED fit yields $T_{\rm eff}=4568_{-11}^{+15}$\,K, $R=0.690_{-0.006}^{+0.006}\,R_{\odot}$ by using the extinction value of $A_{ v}=0$ derived from three-dimensional dust maps \citep{2019ApJ...887...93G}, where $A_{v}$ is the extinction measured in the $V$ band. We find that the source is well fit by a single-star model (see Figure 3), which suggests that the invisible star of J\,0447 is likely a compact object. The astroARIADNE code gives a mass of $M_{\rm K}=0.70 \pm 0.04\,\rm M_{\odot}$ using the stellar evolution model (isochrones). Based on the mass-luminosity relations from \cite{2019ApJ...871...63M}, we obtain the result that the visible star has a mass of $0.68\pm0.02\,\rm M_{\odot}$, which is consistent with the one from the stellar evolution models.

\begin{figure}
\centering
\includegraphics[width=15cm]{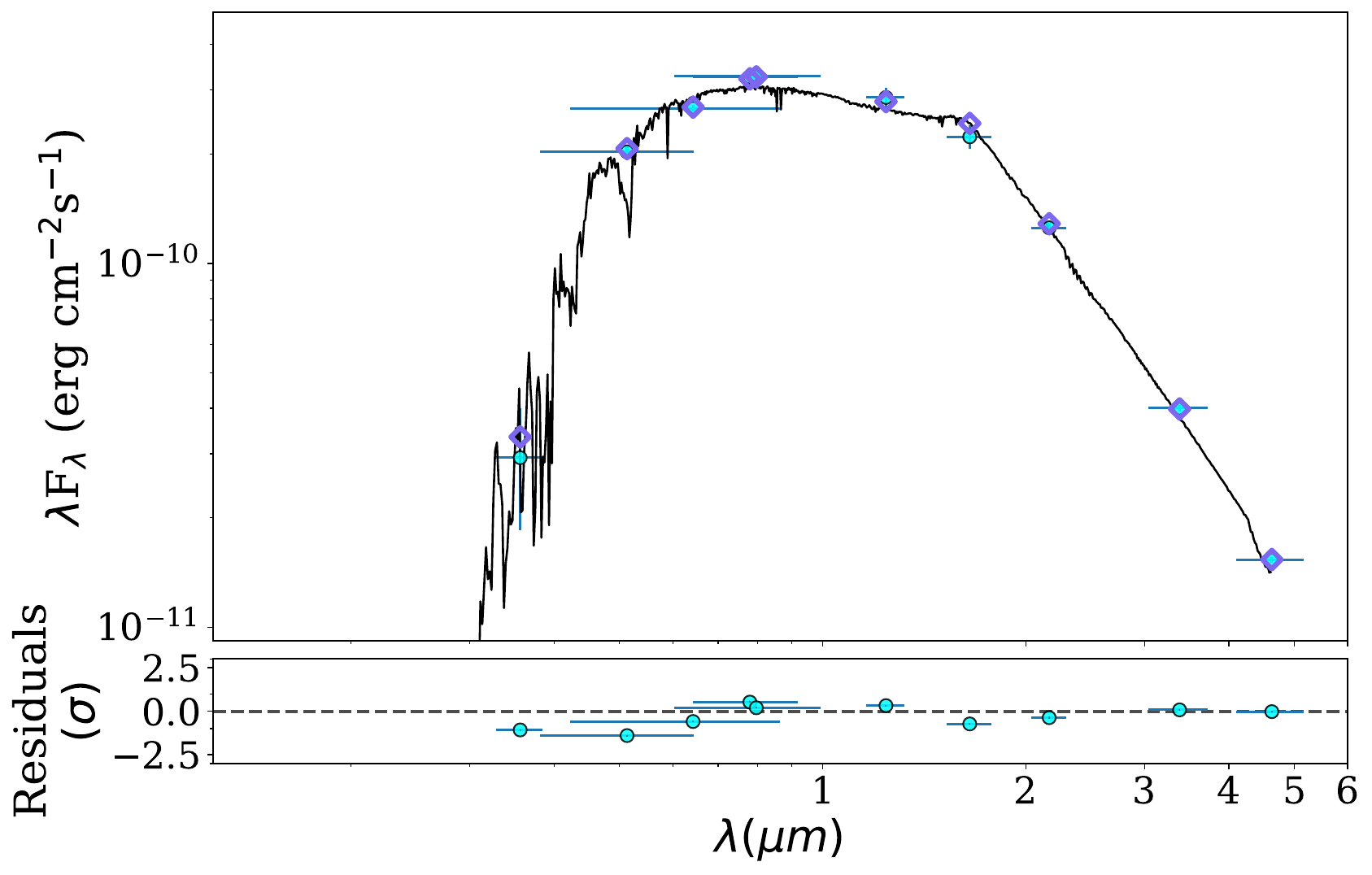}
\caption{The best-fitting SED model for J\,0447 after fixing the extinction parameters $A_{\rm v}=0$. The black curve is the best-fitting model. The green pluses and circles are for the retrieved photometric measurements, while the blue diamonds are for synthetic photometry.}
\label{Halpha}
\end{figure}

\subsection{Optical Spectroscopy}
To further validate the orbital solution of the Gaia DR3, we searched for archived data from other telescopes. The source was also observed by the LAMOST survey. We obtained three low-resolution spectra (LRS) from 2012 December to  2019 January, covering a wavelength range from 369\,nm to 910\,nm \citep{2015RAA....15.1095L}, and 46 medium-resolution spectra (MRS)
from 2017 November to 2018 October, with the blue and red arms covering wavelength ranges from 495\,nm to 535\,nm and from 630\,nm to 680\,nm, respectively. The low- and MRS have spectral resolution $R \sim$ 1800 and $R \sim$ 7500, respectively. 
We derived the barycentric velocity of each spectrum through the cross-correlation technique. In addition, the optical spectroscopy of LAMOST also presents a single-line binary (one visible star only). For four MRS, we use the spectra of the blue arms to measure radial velocities. The cross-correlation has been performed using the spectral template ($T_{\rm eff} = 4568$\,K, log {\it g} = 4.56, [Fe/H] = -0.28) calculated with the MARCS model \citep{2008A&A...486..951G}. The RV was obtained, and we fit our radial-velocity data using the custom Markov Chain Monte Carlo sampler TheJoker \citep{2017ApJ...837...20P}. The best-fitting parameters with 1 $\sigma$ uncertainties are period $P_{\rm orb}=13.3968 \pm 0.00058$\,days, semiamplitude  $K=44.11 \pm 2.57$\,$\rm km$\,$\rm s^{-1}$, eccentricity $e=0.05 \pm 0.02$, and center-of-mass velocity $\gamma = 53.78 \pm 1.37$\,$\rm km$\,$\rm s^{-1}$. This orbital solution also agrees with Gaia DR3.

To derive the surface gravity and metallicity of the K-type star, we use the spectral synthesis codes iSpec \citep{2014A&A...569A.111B,2019MNRAS.486.2075B}. iSpec generates synthetic spectra based on the SME \citep{2017A&A...597A..16P} and MARCS \citep{2008A&A...486..951G} model atmospheres and outputs the best-fitting stellar parameters using a $\chi^2$ minimization process. This fits yield $T_{\rm eff} = 4620 \pm 70 $\,K, $\rm log{\it g} = 4.48\pm 0.17$, and ${\rm [Fe/H]} = -0.41\pm0.07$ (see Figure 5.). 
This result is consistent with the previous SED fitting within the uncertainty. We compared the spectrum of J0447 to the spectra of another star with similar stellar parameters and abundances observed by the LAMOST survey. The two spectra are very similar, which also suggests that the invisible companion of J0447 is a compact object. Notably, the K-type star presents an overabundance of {\it s}-process elements (such as Sr, Ba and La) in LRS from LAMOST using the data-driven Payne method  \citep{2019ApJS..245...34X,2025ApJS..279....5Z}. In addition, \cite{2024ApJ...974...78S} uses the memory-enhanced adaptive spectral network (MEASNet) to search for barium star candidates in the LAMOST low-resolution survey and estimates the abundance of five {\it s}-process elements of the K-type star: [Ba/Fe = 0.86], [Ce/Fe] = 0.50, [Nd/Fe] = 0.22, [Sr/Fe] = 0.28 and [Y/Fe]=0.37 \footnote{doi:10.5281/zenodo.12618383}. Therefore, the K-type star of J0447 is likely a barium dwarf that was contaminated
by an AGB companion. However, high-resolution spectroscopic observations should are required to further confirm the nature of the K-type star. 

\begin{figure}
\centering
\includegraphics[width=15cm]{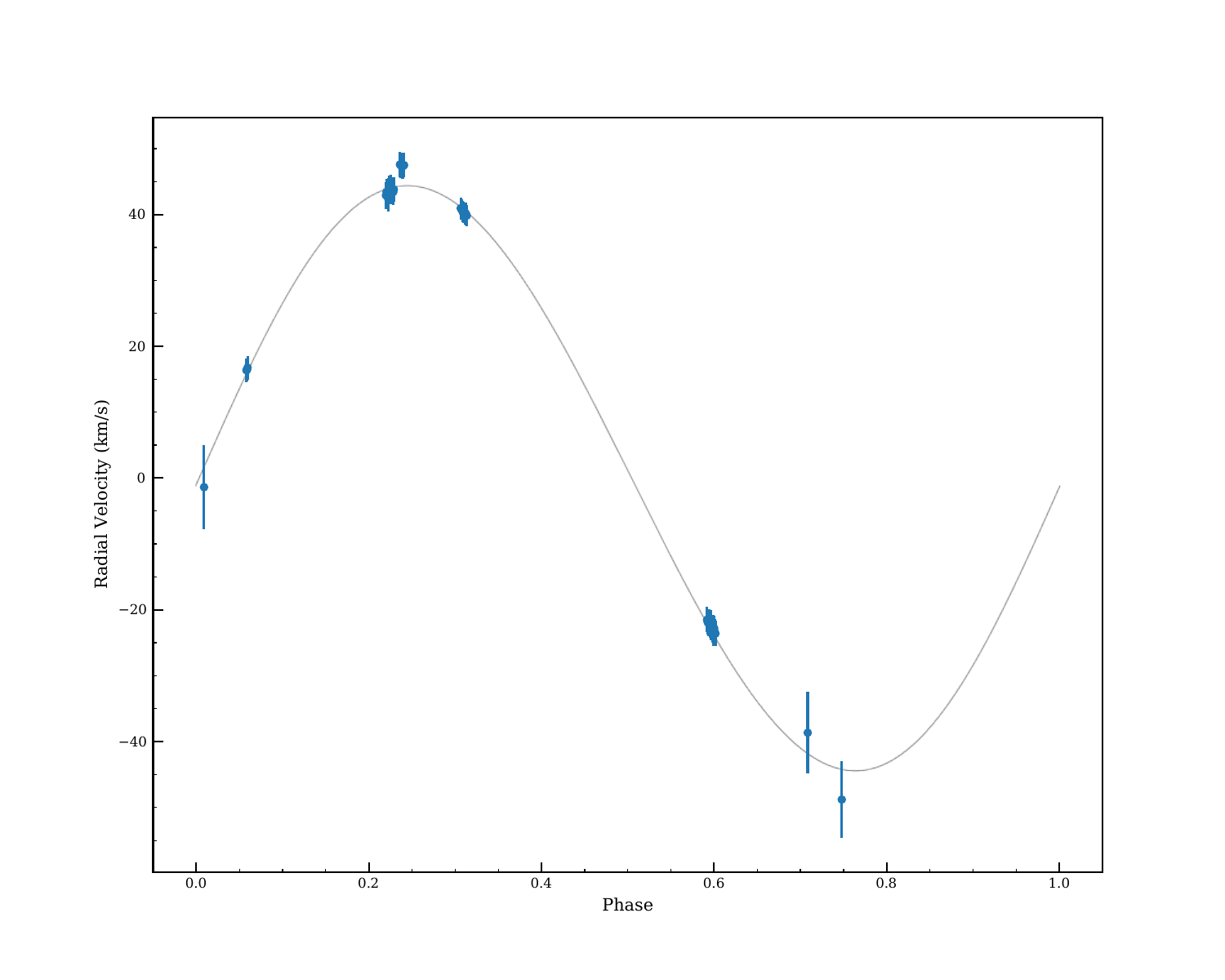}
\caption{The observed RVs of J0447, measured at 49 orbital phases from LAMOST low- to medium-resolution spectra, are plotted as a function of orbital phase with the best-fitting RV curve overlaid.}
\label{rv}
\end{figure}

\begin{figure}
\centering
\includegraphics[width=15cm]{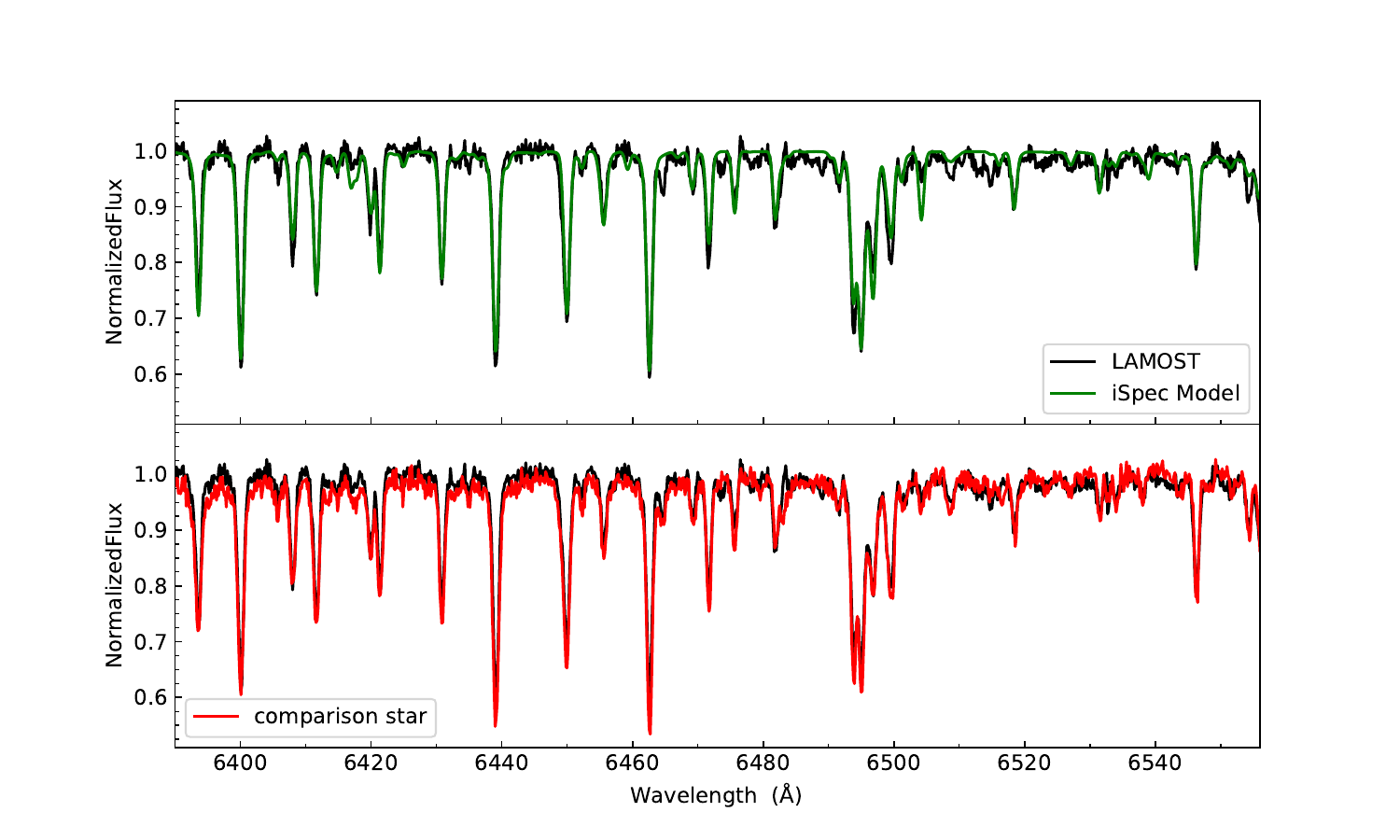}
\caption{The top panel displays a synthetic model spectrum (green) in the red arms, generated using atmospheric parameters derived from iSpec. The bottom panel compares the spectrum of J0447 with that of a comparison star (red) observed by LAMOST, which has similar stellar parameters and chemical abundances. The spectral similarity rules out significant light contamination from a companion.}
\label{RV1}
\end{figure}

\subsection{Radio observation and pulse search}

We observed J\,0447 using FAST on 2024 August 13 and 2025 March 2 (Obs. IDs: PT2024$\_$0156 and PT2024$\_$0133; PI: Jie Lin). The integration times were
1200\,s and 3000\,s, respectively. Observations employed the Pulsar Search Tracking Mode with the central beam of the 19-beam receiver \citep{2020RAA....20...64J}, covering 1.0–1.5\,GHz through 4096 channels. Data were recorded with temporal resolution of 49.152 $\mu$s and full Stokes polarization.

We searched for both periodic signals and single pulses using PRESTO \citep{2001PhDT.......123R} and TransientX \citep{2024A&A...683A.183M}, respectively. For periodic searches, RFI mitigation was performed with RFIFIND provided by PRESTO. The DM search range was 0–30\,pc\,$\rm cm^{-3}$, and the upper limit significantly exceeds the DM value predicted by the YMW16 \citep{2017ApJ...835...29Y} Galactic electron-density model at a distance of 112\,pc. We performed the Fourier-domain acceleration search by adopting zmax=10. No periodic radio pulsation was found in either observation. The DM range for single-pulse searches was identical to that used for the periodic search. No significant single pulse was detected above $S/N$ = 7. By applying the radiometer equation, we estimate a flux density upper limit $S_{\rm min}$ = 2.52\,$\mu$Jy at 1.25 GHz for the 3000 s observation, assuming $S/N_{\rm min}$ = 5 and a 20$\%$ duty cycle.

\subsection{Mass of the unseen companion}

From the resulting RV parameters, we calculate the mass function $f(M)$:
\begin{equation}
	f(M) = \frac{P_{\rm orb}K^{3}(1-e^{2})^{3/2}}{2\pi G}= \frac{M_{\rm C}^{3}{\rm sin}^{3}i}{(M_{\rm C}+M_{\rm K})^2},
\end{equation}
where $M_{\rm C}$ and $M_{\rm K}$ are the masses of the invisible companion and K-type star, respectively. We find $f(M)= 0.12 \pm 0.02\,\rm M_{\odot}$. The mass of the K-type star, $M_{\rm K}=0.70 \pm 0.04 \,\rm M_{\odot}$, was derived from the stellar evolution model. Given the mass function and the mass of the K-type star, the mass of the invisible companion still remains unconstrained due to the lack of inclination $i$. However, the light curve of J0447 does not exhibit significant ellipsoidal variability, preventing constraints on inclination $i$. Therefore, we can only place a lower limit on the invisible companion mass when $i=90^{\circ{}}$. The derived mass for the invisible companion is $M_{\rm C}\,\geq\,0.58 \pm 0.05\,\rm M_{\odot}$. 

The mass of the invisible companion, $M_{C}$, as a function of the inclination is shown in Fig. 6, adopting $M_{\rm K}=0.70\,\rm M_{\odot}$. The integrated probability of inclination $i$ can be calculated as cos $i$ under the assumption of randomly oriented orbital planes. The probability of the invisible companion being a WD, NS or BH is about 82\%, 10\% and 8\%, respectively. Therefore, the invisible companion is likely a WD.

\begin{figure}
\centering
\includegraphics[width=15cm]{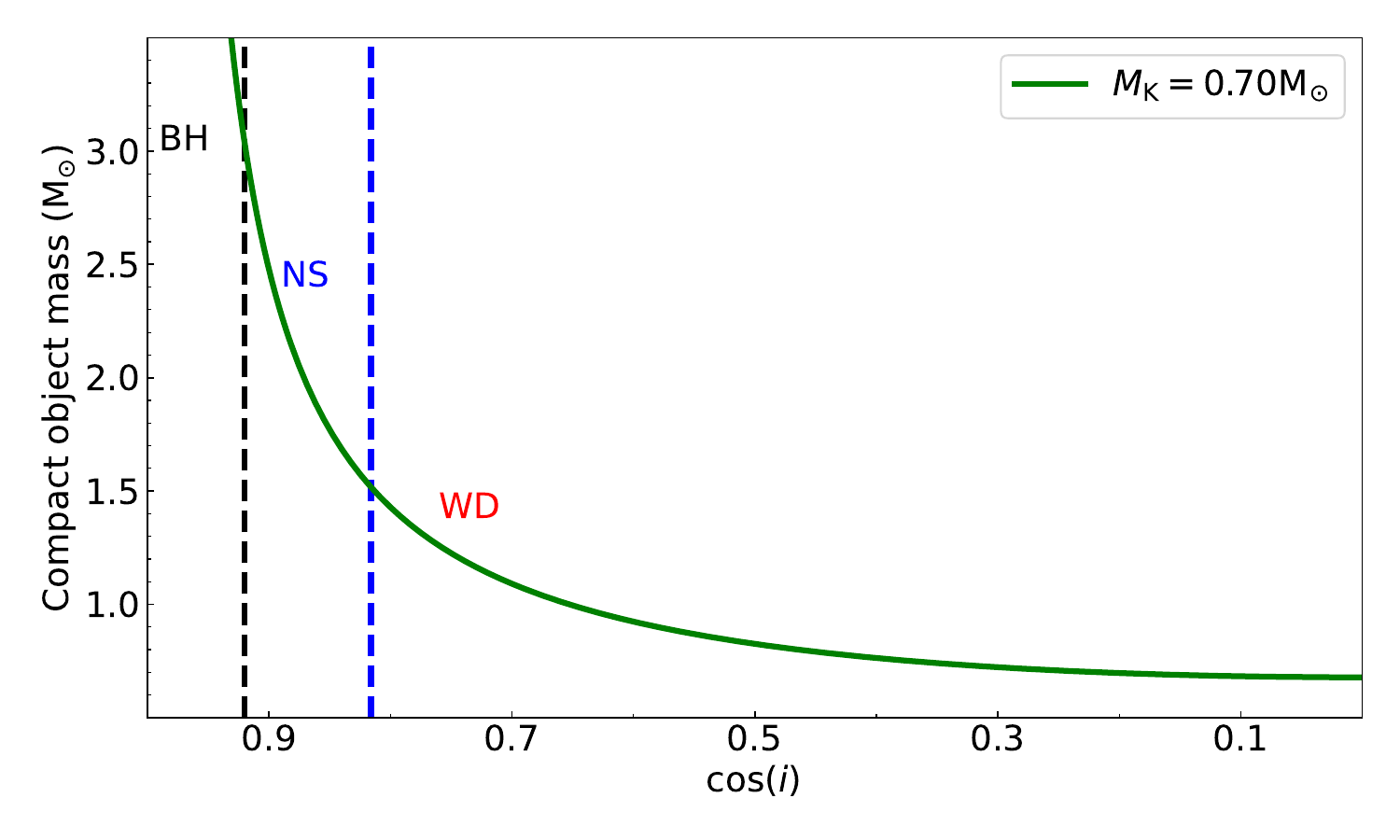}
\caption{Plots of the unseen companion mass as a function of inclination, $i$, on the basis of the measured mass function for 0.7 $\rm M_{\odot}$ K-type star. Assuming that the orbital plane is randomly distributed, the integrated probability of inclination angle can be calculated as cos($i$).}
\label{massp}
\end{figure}

\section{Discussion}

\subsection{MS companion}

The luminosity of the MS companion with a mass of $M_{\rm C}\,\geq\,0.58 \pm 0.05\,\rm M_{\odot}$ is almost consistent with that of the K-type star. The MRS from LAMOST has already covered the semiamplitude phase of the RV curve. Therefore, in this case, the LAMOST spectra are sufficient to reveal two sets of spectral lines and the variations in the composite line profiles with orbital phase. In addition, the spectra of J0447 are well fitted by single-star models and are consistent with those of single stars observed by LAMOST. Therefore, the presence of an MS can be definitively ruled out.

\subsection{Neutron star or black hole}
Due to the lack of constraints on the orbital inclination, only the minimum mass of the unseen companion can be determined. Therefore, we also consider the possibility that the unseen companion is an NS nor a BH, particularly given that this compact binary system resides as close as $\sim$ 100\,pc from Earth. Over the past two decades, many experiments have detected pulses of radionuclide $^{60} \rm Fe$ in deep-sea deposits between 2 and 3\,Myr ago \citep{1999PhRvL..83...18K,2016Natur.532...69W}. In addition, there are measurements of $^{60} \rm Fe$ in the lunar
regolith \citep{2016PhRvL.116o1104F}, in cosmic rays \citep{2016Sci...352..677B}, and in Antarctic snow \citep{2019PhRvL.123g2701K}. The discovery of these live radioactive isotopes is believed to be recent astrophysical explosions such as core-collapse supernovae (CCSN) within $\sim$ 100\,pc of Earth. If the unseen companion is indeed an NS or BH, it will be the nearest NS or BH. Furthermore, we performed a kinematic analysis of J\,0447's orbit in the galaxy using the Gaia astrometric solution and the systematic RV and found that it can pass through our solar neighborhood spanning the past 2-3\,Myr (see Figure 7.). This may suggest that the radionuclide $^{60} \rm Fe$ signal is associated with its CCSN event.

Either NSs or BHs are expected to receive natal kicks that impart eccentricity to their orbits. In the specific case where the natal kick arises solely from spherically symmetric mass loss and the presupernova orbit is circular \citep{1961BAN....15..265B}, the resulting eccentricity e is given by 
\begin{equation}
	e= \frac{ \Delta M }{M_{\rm c}+M_{2}},
\end{equation}
where $\Delta M$ is the mass lost during the supernova, $M_{\rm c}$ is the mass of the resultant compact object (NS or BH), and $M_{2}$ is the mass of the companion star. For an NS formed in this scenario, we consider a 8 $\rm M_{\odot}$ star undergoing an ultra-stripped core-collapse SN with $\sim$\,0.3 $\rm M_{\odot}$ ejecta \citep{2018Sci...362..201D, 2020ApJ...900...46Y}, forming a typical 1.4 $\rm M_{\odot}$ NS around a 0.7 $\rm M_{\odot}$ MS companion. The resulting eccentricity (e\,$\sim$\,0.14) is significantly larger than the system of J0447. Asymmetric supernovae can impart strong natal kicks to NSs, generating even higher eccentricities than those produced by spherically symmetric mass loss alone. Consequently, the highly circular orbit of J0447 is difficult to reconcile with the mechanisms of NS formation. Furthermore, deep FAST radio observations reveal no radio pulsed signals when ignoring unfavorable beaming geometries. This further strengthens the evidence that the unseen companion is unlikely to be an NS.

On the other hand, if the unseen companion is a BH and the natal kick was purely due to mass loss during its formation, the orbital eccentricity (e $\sim$ 0.05) of the system can place a constraint on the mass of the BH. Assuming a minimum mass loss of $\sim$\,0.3 $\rm M_{\odot}$, a BH mass of $M_{\rm BH}$\,$\geq$\,$5.3\,\rm M_{\odot}$ would be required to fully account for the observed eccentricity. For a random distribution of the orbital inclination angle, the probability of obtaining such a low inclination angle ($\sim$ $18^{\circ{}}$) is less than 5\% since the BH mass exceeds $5.3\,\rm M_{\odot}$. Furthermore, if the SN is asymmetric, imparting an additional kick on the BH, this would likely result in a higher BH mass and a concomitantly lower probability of having a low inclination angle, as the kick contributes additional eccentricity to the system. Although the possibility that the unseen companion is an NS or BH cannot be entirely ruled out, it appears highly unlikely. Therefore, the unseen companion is likely a WD, and we proceed under this assumption.

\begin{figure}
\centering
\includegraphics[width=15cm]{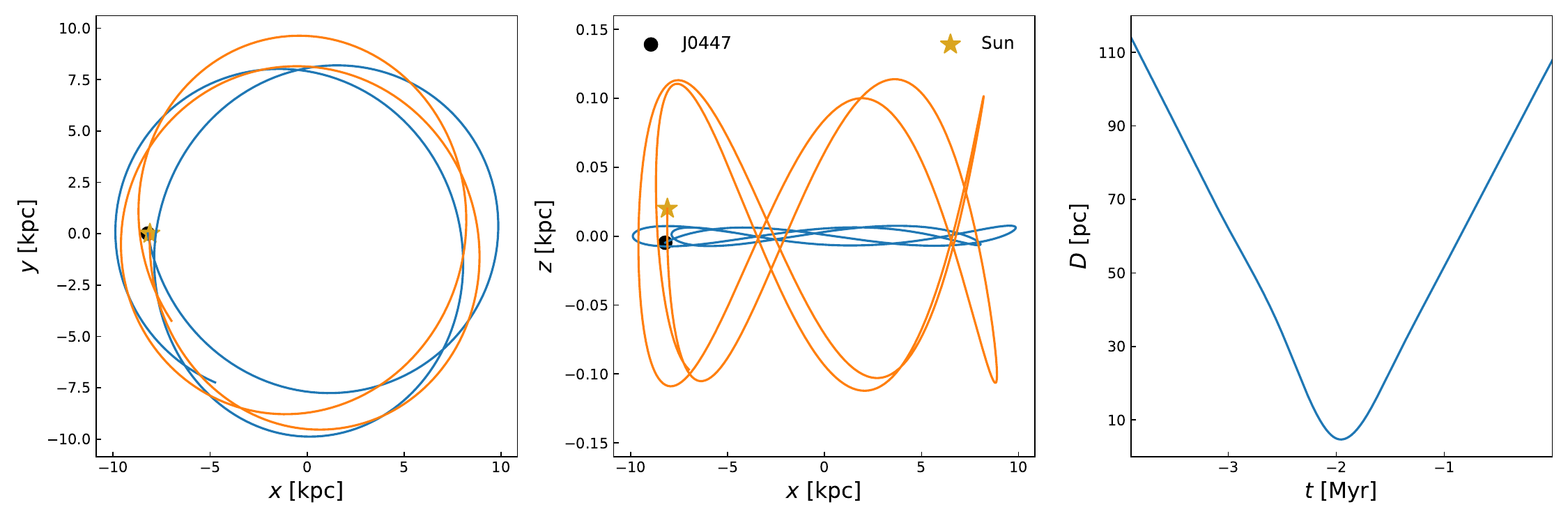}
\caption{Left-hand and middle panels show the  Galactic $xy$-plane and $xz$-plane orbital projections of J0447 and the Sun, respectively, calculated via 500 Myr backward integration from the measured proper motion and center-of-mass RV \citep{2017JOSS....2..388P}. The right-hand panel shows the temporal evolution of the distance between J0447 and the Sun over the past several Myrs. }
\label{xyrt}
\end{figure}

\subsection{A white dwarf in postcommon envelope binaries}
We identified a close binary star consisting of a K-type star and most likely a WD. Only a minimum mass of the unseen companion is obtained due to the lack of inclination angle. In what follows, we compare the properties of our targets with other related
classes of binaries and investigate possible implications for WD binary formation and evolution by reconstructing their past.

\subsubsection{A population of PCEBs in the literature}
The SDSS efficiently identifies substantial populations of close white dwarf–main-sequence (WDMS) binaries \citep{2009ApJS..182..543A,2007MNRAS.382.1377R}. \cite{2010A&A...520A..86Z} compiled a sample of 60 PECBs, comprising 35 systems newly identified through the SDSS and 25 previously known systems. In addition, the ``white dwarf binary pathways survey" also identified some PCEBs that consist of a WD plus an intermediate mass companion star of spectral type AFGK \citep{2021MNRAS.501.1677H,Hernandez2022a,Hernandez2022b}. \cite{2026arXiv260100439S} identified 39 detached eclipsing WDMS
binaries using ZTF light curves. With the exception of the IK Peg system ($\sim$ 21.7 days), the orbital periods of all other systems are less than 10 days. Beyond IK Peg, multiple long-period WDMS self-lensing binaries (SLBs) have been identified through systematic pulse surveys in the Kepler light curves \citep{2014Sci...344..275K,2018AJ....155..144K}. More recently, \cite{2024MNRAS.52711719Y} identified five PECBs hosting massive WD candidates paired with MS companions of spectral types earlier than M. These systems exhibit long orbital periods (18-49 days) derived from spectroscopic and astrometric data - classified as SB1 or astrometric + spectroscopic (AstroSpectroSB1) in the Gaia DR3. As illustrated in Figure 8, only three self-lensing binaries (SLBs) reside near the $P$-$M_{\rm wd}$ relation for WDs, whereas the remaining sample systems deviate significantly from this trend \citep{2011ApJ...732...70L}. This deviation indicates that the outliers likely formed through common-envelope evolution (CEE) rather than stable Roche-lobe overflow (RLOF). The three SLBs exhibit long orbital periods and near-circular orbits, consistent with formation via stable mass transfer from an evolving primary - a mechanism analogous to that observed in field and open-cluster blue stragglers \citep{2018AJ....155..144K}.

\begin{figure}
\centering
\includegraphics[width=15cm]{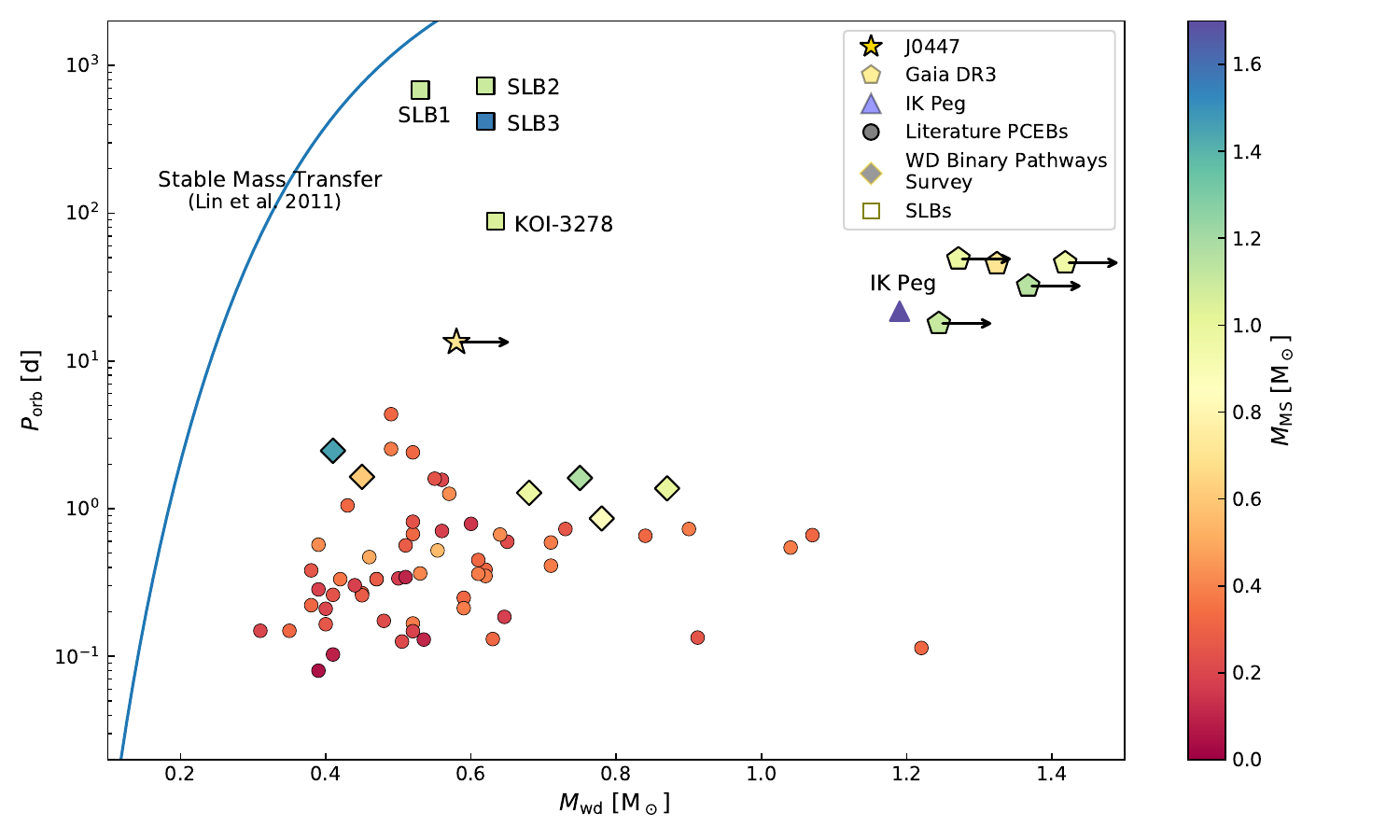}
\caption{Orbital period versus WD mass distribution of PCEBs, including the J0447 system presented in this work. The blue line shows the theoretical $M_{\rm wd}$-$P_{\rm orb}$ relation for the stable mass transfer case by \cite{2011ApJ...732...70L}. A large of number PCEBs with M-dwarf companion (circle markers) compiled by \cite{2010A&A...520A..86Z} exhibit an orbital period distribution concentrated below 10\,days. The six PCEB systems from the ‘White Dwarf Binary Pathways Survey’ are denoted by diamond markers \citep{2021MNRAS.501.1677H,Hernandez2022a,Hernandez2022b}. The pentagon and triangle markers denote five PCEBs hosting massive white dwarf binary candidates from Gaia DR3 \citep{2024MNRAS.52711719Y} and IK Peg, respectively. These systems are distinguished from other known PCEBs by their long orbital periods. The KIC 8145441 system \citep{2014Sci...344..275K} and other SLBs \citep{2018AJ....155..144K}, indicated by square markers, were all detected by Kepler. }
\label{pceb}
\end{figure}

For PCEBs with orbital periods less than 10 days, \cite{2010A&A...520A..86Z} demonstrated that all known systems can be reconstructed using a relatively low CE efficiency ($\alpha_{\rm CE}$ $\sim$ 0.25), indicating inefficient orbital energy transfer during envelope ejection. In contrast, there are currently several PCEB systems with long orbital periods that range from a few tens to a few hundreds of days. It has been claimed that these systems cannot be explained by CEE without contributions from recombination energy. However, \cite{2024A&A...687A..12B} argued that extra energy is not required to explain PCEB systems with long orbital periods when the WD progenitor has to be highly evolved TP-AGB star at the onset of the CE evolution. However, it remains to be observationally tested whether all wide-orbit systems necessarily evolve into highly evolved TP-AGB stars before CE evolution. If so, the heavy {\it s}-process elements, such as Ba, La, or Ce, and light {\it s}-process elements, such as Sr, Y, or Zr, might be detectable in the spectra of some companion stars \citep{2023A&A...671A..97E}. Interestingly, the LRS from LAMOST indicates potential {\it s}-process element enrichment in the K-type star of our target, suggesting that the WD progenitor likely went through the TP-AGB phase.

\subsubsection{The evolutionary history of J0447}

From the previous section, we know that the MS donor is populated by the {\it s}-process elements, which are produced from TP-AGB star. This means that the progenitors of the WDs should undergo TP-AGB phases. Assuming that the compact object is a WD, the formation scenario of the J0447 system is as follows. In an MS binary, the primary evolves to the TP-AGB phase and fills its Roche lobe, while the secondary is still an MS star. Then the system experiences an unstable mass transfer and enters the common envelope phase. After the ejection of the CE, the system evolves into a binary system consisting of a WD and an MS star. Below we try to constrain the binary parameters of the progenitors of J0447.

For a binary system to enter a CE phase, the mass ratio must exceed a critical mass ratio. In this case,  we refer to the critical mass ratio given by \citet{2002MNRAS.329..897H} (see their Eq. 57).
For the CE process, we adopt the energy budget prescription, i.e. 
\begin{equation}
\alpha_{\rm CE}(\frac{GM_{\rm C}M_{\rm K}}{a_{\rm f}}-\frac{GM_{\rm i}M_{\rm K}}{a_{\rm i}}) = |E_{\rm bind}|,
\end{equation}

\begin{equation}
E_{\mathrm{bind}}=\int_{M_{\mathrm{c}}}^{M_{\mathrm{i}}} (-\frac{G m}{r} +\varepsilon_{\mathrm{int }})\;\mathrm{d} m,
\end{equation}
where $G$ is the gravitational constant; $M_{\rm C}$ and $M_{\rm K}$ are the masses of the compact object and companion star, respectively; $a_{\rm i}$ and $a_{\rm f}$ are the binary separation at the onset of CE and after the ejection of CE, $M_{\rm i}$ is the progenitor mass of the compact object, $M_{\rm c}$ is its core mass at the onset of CE and $\alpha_{\rm CE}$ is the CE efficiency. 
$E_{\rm bind}$ is the binding energy of the envelope of the progenitor of the WD. $m$ is the mass coordinate, $r$ is the radius and $\varepsilon_{\rm int}$ is the internal energy, which includes the radiation energy and thermal energy, but not the recombination energies.

\begin{figure}
    \centering
    \includegraphics[width=0.45\linewidth]{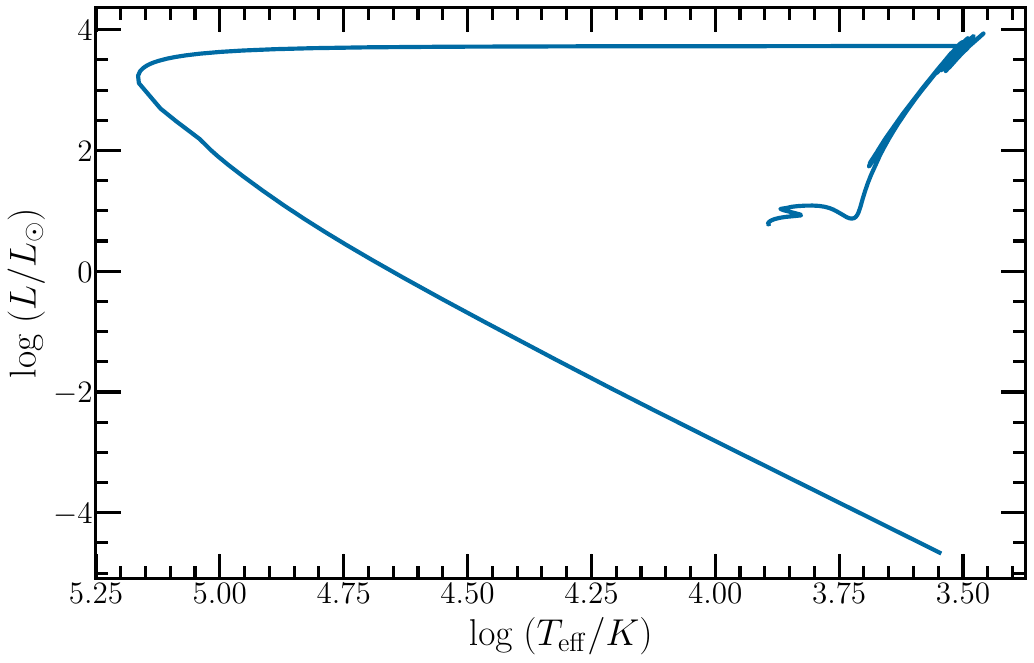}
    \includegraphics[width=0.45\linewidth]{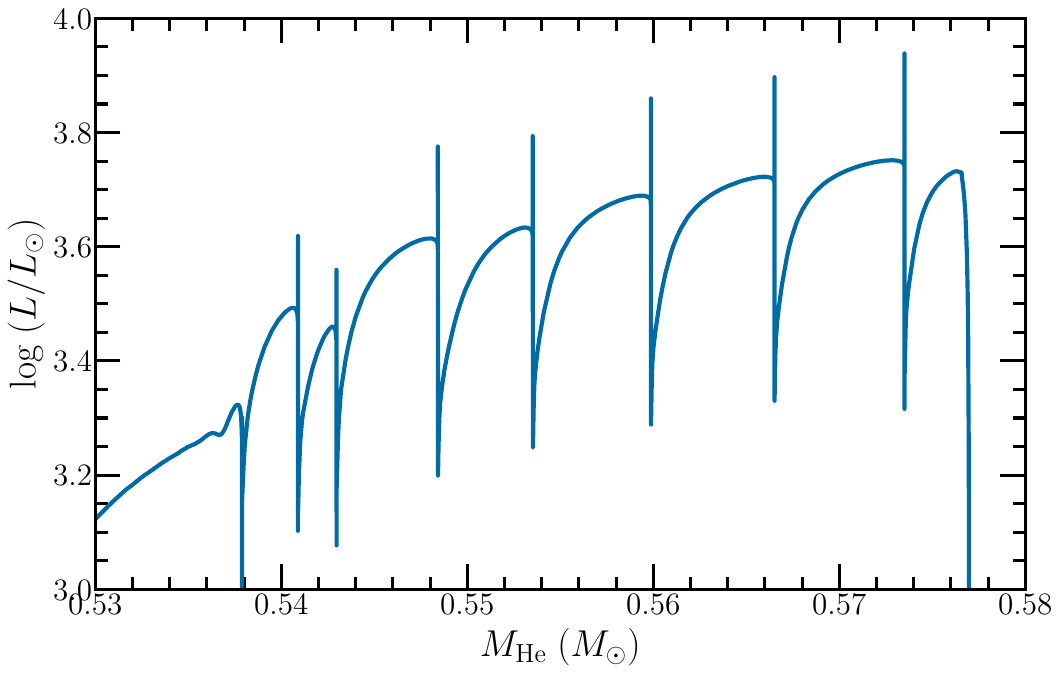}
    \caption{Left panel: evolution of a $1.5\;M_{\odot}$ star in the HR diagram; Right panel: evolution of luminosity as a function of He core mass around the TP-AGB phase for a $1.5\;M_{\odot}$ star.}
    \label{fig:agbex}
\end{figure}

In order to obtain the binding energy of the progenitors of WDs, we compute a grid of single stellar evolution models with the stellar evolution code Modules for Experiments in Stellar Astrophysics (\textsc{MESA}, \citealt{2011ApJS..192....3P,2013ApJS..208....4P,2015ApJS..220...15P,2018ApJS..234...34P,2019ApJS..243...10P}). The initial stellar mass ranges from $1.0\;M_{\odot}$ to $8\;M_{\odot}$ with a step of $0.50\;M_{\odot}$. 
The initial chemical abundances are assumed to be $X$ = 0.73, $Y$ = 0.26, $Z$ = 0.01. The mixing length parameter is assumed to be 2.0 and overshooting is not considered in our calculation.  
Regarding the stellar wind, we adopt the prescription of \citet{1975MSRSL...8..369R} with a wind efficiency parameter of 0.5 for RG stars and the prescription of \citet{1995A&A...297..727B} with a wind efficiency parameter of 0.10 for AGB stars. The stars are evolved from the zero-age MS to the WD phase. 
In Fig.~\ref{fig:agbex}, we show an example of the single stellar evolution track in our calculation. The initial mass in that plot is $1.5\;M_{\odot}$.

\begin{figure}
    \centering
    \includegraphics[width=0.45\linewidth]{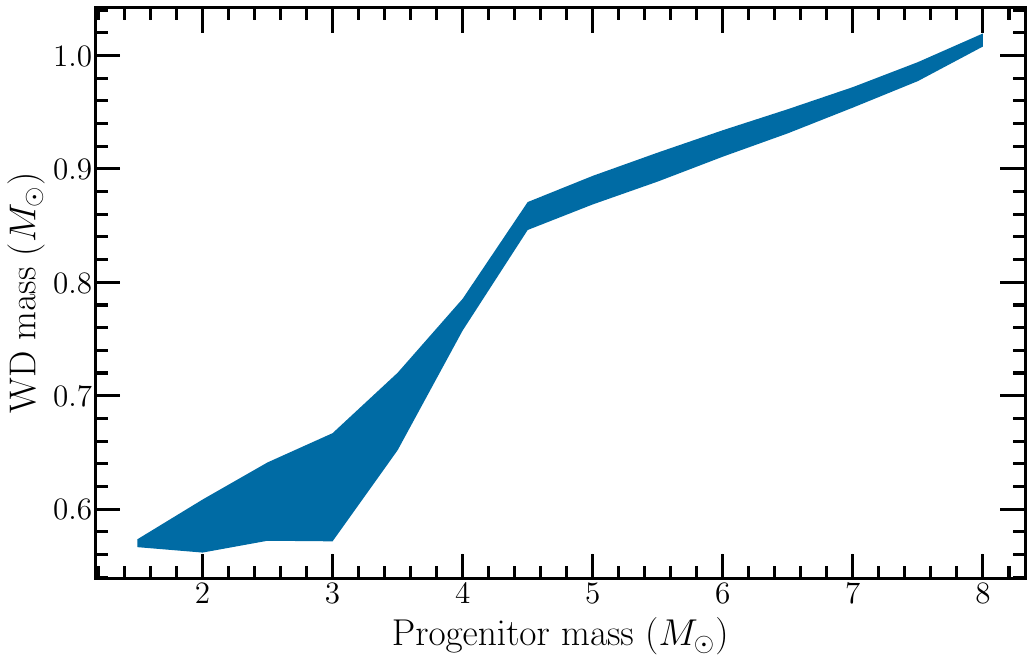}
    \includegraphics[width=0.45\linewidth]{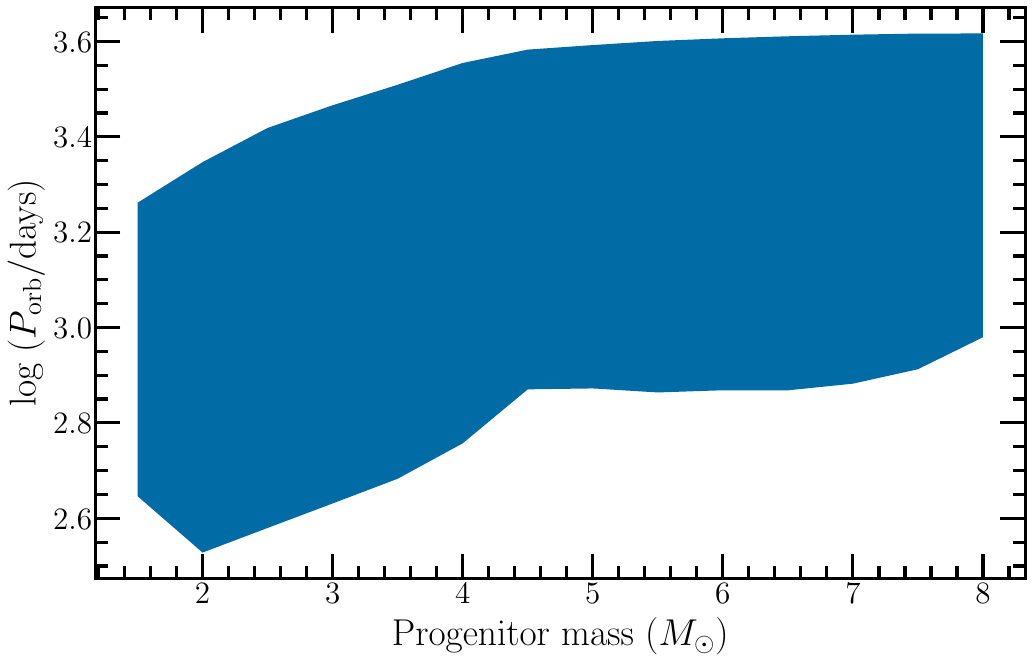}
    \caption{The initial binary parameter space of the progenitors of J0447 in the WD mass versus its progenitor mass panel and progenitor mass versus initial orbital period panel.}
    \label{fig:m_mwd}
\end{figure}

We constrain the WD mass to be within the range of $0.56-1.40\,\rm M_{\odot}$ and $\alpha_{\rm CE}$ within 0-1.0. Then we can constrain the progenitor mass of the WD and also the initial orbital period, which are shown in Fig.~\ref{fig:m_mwd}. From this plot, we can find that the progenitor mass of the WD is around $1.5-8.0 M_{\odot}$ and the initial orbital period is around $340-4200$ days.


%
%

\section{Conclusions}
We report an SB1 that contains a K-type star and a compact object in the solar neighborhood. Although the nature of a compact object has not yet been determined, this binary system is likely a PCEB consisting of WD and K-type companions with intermediate orbital periods between tight and wide orbital periods. If the compact object is a BH, it will be the nearest BH, which may help us to understand the origin of radionuclide elements in the solar neighborhood. However, analyses based on machine learning techniques applied to LRS indicate a potential overabundance of {\it s}-process elements in the K-type star, which further supports that the compact object is a WD. Furthermore, considering the relatively low inclination probability presented in section 4.2, an NS or BH companion is a less likely scenario. For a PCEB with intermediate orbital periods, we found that extra energy is not required to explain this system when the WD progenitors experienced a highly evolved TP-AGB phase at the onset of the CE evolution. 
Importantly, the K-type companion, a potential Barium star candidate, shows an overabundance of heavy elements heavier than iron produced by the {\it s}-process.  This is because the material formed in the interior of the more evolved WD progenitor is being transferred onto the companion star. This system provides insight into the efficiency of CE ejection and suggests that additional energy sources are not required to expel the envelope during the CEE. Nevertheless, high-resolution spectroscopy observations are crucial to verify the {\it s}-process overabundance in the K-type companion and to test for a possible misclassification as a barium star. If the K-type companion star is confirmed to be a barium star, this would represent the shortest orbital period barium star binary system discovered to date, which may hold significant implications for understanding the formation and evolution of barium star binary systems.


\acknowledgments 
We thank Dr. Weimin Gu, Dr. Smith, D. A, and Dr. Lin Lan for their helpful discussions. The Guoshoujing Telescope (the Large Sky Area Multi-Object Fiber Spectroscopic Telescope, LAMOST) is a National Major Scientific Project built by the Chinese Academy of Sciences. Funding for the project has been provided by the National Development and Reform Commission. LAMOST is operated and managed by the National Astronomical Observatories, Chinese Academy of Sciences. FAST is a Chinese national mega-science facility, operated by the National Astronomical Observatories, Chinese Academy of Sciences. ZTF is a public-private partnership, with equal support from the ZTF Partnership and from the U.S. National Science Foundation through the Mid-Scale Innovations Program (MSIP). J.L. acknowledges the support of the Postdoctoral Fellowship Program of CPSF under Grant Number GZC20240905. Y.~L. acknowledges the support of the Boya fellowship of Peking University and the China Postdoctoral Science Foundation (No. 2025T180924). H.~L. acknowledges the support of the CAS "Light of West China", the Young Talent Project of Yunnan Revitalization Talent Support Program, the Yunnan Fundamental Research Project (No. 202401BC070007).
This work is supported by the National Natural Science Foundation of China under grant 12403055, 12221003, 12335009, 12288102, 12333008 and 12422305.




\end{document}